\documentclass[epj-spec]{svjour}
\usepackage{graphicx}

\def\veck{\ensuremath{\mathbf{k}}}
\def\vecp{\ensuremath{\mathbf{p}}}
\def\vecv{\ensuremath{\mathbf{v}}}
\def\vecg{\ensuremath{\mathbf{g}}}
\def\vecDeltav{\ensuremath{\mathbf{\Delta v}}}
\def\ket#1{\ensuremath{\left|#1\right>}}

\def\dt{\mathrm dt}
\def\TR{{\ensuremath{T_\mathrm{R}}}}

\usepackage{setspace}
\begin{document}
\title{Atom interferometry based on light pulses : application to the high precision measurement of the ratio $h/m$ and the determination
of the fine structure constant}

\author{ M.~Cadoret\inst{1}, E. De
Mirandes\inst{1}, P. Clad\'e\inst{1}, F. Nez\inst{1}, L.
Julien\inst{1}, F. Biraben\inst{1} and
S.~Guellati-Kh\'elifa\inst{1}\fnmsep\inst{2}\fnmsep\thanks{\email{guellati@spectro.jussieu.fr}}
}
\institute{Laboratoire Kastler Brossel, Universit\'e Pierre et Marie
Curie, ENS, CNRS, 4 place Jussieu, 75252 Paris Cedex 05, France\and
LNE-INM, Conservatoire National des Arts et M\'etiers, 61 rue du
Landy, 93210 La Plaine Saint Denis, France}
\abstract{In this paper we present a short overview of atom
interferometry based on light pulses. We discuss different
implementations and their applications for high precision
measurements. We will focus on the determination of the ratio $h/m$
of the Planck constant to an atomic mass.  The measurement of this
quantity is performed by combining Bloch oscillations of atoms in a
moving optical lattice with a Ramsey-Bord\'e interferometer.
} 
\maketitle
\section*{Introduction}
\label{intro}  Intense progress in laser cooling has led to the
building of reliable atom interferometers. A broad overview of
accomplishments in matter wave interferometry was presented by
Berman over a decade ago \cite{Berman}. To implement such an
interferometer, optical elements based both on the mechanical forces
of light \cite{Gould86,Borde89,Riehle91,KasPRL91,Ster92,Jacquey07}
and nanofabricated structures \cite{Kieth88,Carnal91,Shimizu92} have
been investigated. Right from the outset, light-based atom
interferometry became a technique of choice for high precision
measurements. This method has benefited greatly from the impressive
progress in laser technology. In this lecture we focus on atom
interferometers based on light pulses. In such an interferometer we
consider the atomic interference of internal states.  The spatial
separation of the atoms is produced by the momentum recoil induced
by the electromagnetic field used to drive the atoms from one
internal state to another. In this way, the internal and external
degrees of freedom are strongly linked. Conceptually, such an
interferometer provides a measurement of the recoil
frequency. This lecture is divided in two main parts. The first is
devoted to the analysis of atom interferometer based on light
pulses. After a description of the basic concepts such as Rabi
oscillations and stimulated Raman transitions, we use a simple
formalism based on a plane wave to describe Ramsey-Bord\'e
interferometers. We discuss different configurations and their
application to high-precision measurements. In the second part, we
describe an experiment underway in our group at Laboratoire Kastler
Brossel which combines Bloch oscillations with a Ramsey Bord\'e
interferometer. Using this approach, we aim to determine the ratio
$h/m_{\mathrm{Rb}}$ (where $m_{\mathrm{Rb}}$ is the mass of the
rubidium atom) and hence a value for the fine structure constant
$\alpha$.

\section*{\large{Part I: Atom interferometry based on optical pulses} }

\section{Basic concepts}
\subsection{Two-level atom interacting with travelling waves: Rabi oscillations}
\label{sec:2} Let us consider a  two-level atomic system
($|g\rangle$ and $|e\rangle$) with respective energies $\hbar
\omega_g$ and $\hbar \omega_e$. The levels are coupled by an
electromagnetic field of angular frequency $\omega$ and wave vector
$\mathbf{k}$. We denote by $\Omega$ the Rabi frequency which
represents the coupling constant. In the rotating wave
approximation, the Hamiltonian is given by:
\begin{equation}
H=\frac{\mathbf{\widehat{p}}^2}{2m}+\hbar \omega_e |e\rangle \langle
e|+\hbar \omega_g |g\rangle\langle g|+\left[\hbar \Omega
e^{i(\mathbf{k}\hat{\mathbf{x}}-\omega t)}|e\rangle\langle g|+\hbar
\Omega^* e^{-i(\mathbf{k}\hat{\mathbf{x}}-\omega t)}|g\rangle\langle
e|\right]
\end{equation}
where $\mathbf{\widehat{p}}$ operates on the momentum part of the
atomic state.

A complete description of the atom-light interaction should include
both the internal energy states and the external degrees of freedom.
Such a description is necessary to analyze an atom interferometer.
One should consider explicitly the propagation of spatial wave
packets in order to calculate the phase shifts between interfering
paths due both to the free-space propagation and to the atom-light
interaction. The usual approach is to consider the atomic wave
packets as the sum of plane-wave momentum states. The momentum
transfer due to the interaction of the electromagnetic field is
calculated for a given  plane-wave component, after which the
integral over all of the momentum states in the atomic ensemble is
performed.  The atomic states are labeled  by $|g,
\mathbf{p_g}\rangle$ or  $|e, \mathbf{p_e}\rangle$ ($|g,
\mathbf{p_g}\rangle=|g\rangle \bigotimes |\mathbf{p_g}\rangle$ and $|e,
\mathbf{p_e}\rangle=|e\rangle \bigotimes |\mathbf{p_e}\rangle$). In the momentum
basis, the spatial dependence arises via the translation operator
$e^{\i\mathbf{k\cdot \hat{\mathbf{x}}}}$:

\begin{equation}
e^{i\mathbf{k\cdot \hat{\mathbf{x}}}}|\mathbf{p}\rangle =|\mathbf{p+\hbar k}\rangle
\end{equation}
This equation expresses the well-known result whereby the absorption
of a photon of wave vector $\mathbf{k}$ changes the atomic momentum
by $\hbar \mathbf{k}$. The light field couples the quantum states
$|g, \mathbf{p}\rangle$ and $|e, \mathbf{p+\hbar k}\rangle$.

For simplicity, we use the transformations $|\tilde{g},\rangle = e^{i\omega_g t}|g\rangle$ and
$|\tilde{e}\rangle = e^{i\omega_e t}|e\rangle$. The hamiltonian becomes:

\begin{equation}
H=\frac{\mathbf{\widehat{p}}^2}{2m}+\left[\hbar \Omega
e^{i(\mathbf{k}\hat{x}-\phi(t))}|\tilde{e}\rangle\langle\tilde{g}|+\hbar \Omega^*
e^{-i(\mathbf{k}\hat{x}-\phi(t))}|\tilde{g}
\rangle\langle \tilde{e}|\right]
\end{equation}
where the phase $\phi(t) = \delta t$ , with detuning $\delta =
\omega - (\omega_e-\omega_g)$. This hamiltonian describes the
transition from $|g\rangle$ to $|e\rangle$ associated with the
transfer of photon momentum $\hbar \mathbf{k}$.

The time evolution of the quantum state $|\psi(t)\rangle$ can be
expressed in terms of the time-dependent coefficients
$a_{g,\mathbf{p}}(t)$ and $a_{e,\mathbf{p}+\hbar \mathbf{k}}(t)$:

\begin{equation}
|\Psi (t)\rangle = a_{e,\mathbf{p}+\hbar
\mathbf{k}}(t)|\tilde{e},\mathbf{p}+\hbar \mathbf{k}\rangle e^{-i
\frac{|\mathbf{p}+\hbar \mathbf{k}|^2}{2m\hbar}
t}+a_{g,\mathbf{p}}(t)|\tilde{g}, \mathbf{p}\rangle e^{-i
\frac{|\mathbf{p}|^2}{2m\hbar}t}
\end{equation}
These coefficients can be calculated by solving the Schr\"{o}dinger
equation :

\begin{equation}
i\hbar \frac{d}{dt}|\Psi (t)\rangle = H |\Psi (t)\rangle
\end{equation}
In the limit where $\phi(t)$ is constant during the pulse,
 the probability of finding an atom in state $|e\rangle$ after an
interaction time $\tau$ is then given by :
\begin{equation}
|a_{e,\mathbf{p}+\hbar \mathbf{k}}(t)|^2=
\frac{1}{2}\left(1-\cos\Omega\tau\right)
\end{equation}
The atom undergoes the well-known Rabi oscillations between the two
states $|g\rangle$ and $|e\rangle$.

Two configurations are relevant for atom interferometry in the
particular case where $a_g(t_0)=1$ and $a_e(t_0)=0$ :

\begin{itemize}
\item \textbf{\textit{Case I }}: $\pi/2$-light pulse here corresponding to $\Omega\tau=\pi/2$
\begin{equation}
|\Psi (t)\rangle = \frac{1}{\sqrt{2}}\left[|\tilde{g},
\mathbf{p}\rangle-i|\tilde{e}, \mathbf{p+\hbar k}\rangle
e^{-i\phi(t)}\right] \label{PI/2}
\end{equation}
\\
\item \textbf{\textit{Case II }}: $\pi$-light pulse \textit{i.e.} $\Omega\tau=\pi$

\begin{equation}
|\Psi (t)\rangle = - i |\tilde{e}, \mathbf{p+\hbar k}\rangle
\label{PI}
\end{equation}
\end{itemize}

When the atom is subjected to a $\pi/2$ light pulse, the photon of
momentum $\hbar\mathbf{k}$ puts it into a coherent and equal
superposition of both energy states. The recoil imparted to the atom
gives rise to two coherent wave packets (in each internal state)
separated by a velocity $v_r=\hbar\mathbf{k}/m$. In others words,
this $\pi/2$ light pulse plays a role analogous to the 50-50 beam
splitter in classical optics. In the same way, when the coupling
parameters are chosen such that $\Omega_{eg} \tau=\pi$ (a $\pi$-pulse), the probability of finding an atom in $|e,
p+\hbar\mathbf{k}\rangle$ after a time $\tau$ is equal to unity; in
this case the light pulse behaves like a mirror.

We should emphasize that two terms play an important role for
calculating phase shift in atom interferometer: the kinetic energy
$p^2/2m$ (external degree of freedom) and the phase $\phi(t)$. The
latter depends both on the laser frequency and on the internal
energy of the atoms. In this particular case, we have chosen a fixed
laser frequency and constant internal energy for the atom. More
generally, to include the situations where the detuning $\delta$ varies through the laser frequency (frequency
sweep or phase jump, for example) or through the energy of the atomic levels (\textit{i.e.} light shift, Zeeman effect),
 one has to evaluate $\phi(t) = \int \delta(t) \dt$.

Up to now we have neglected spontaneous emission. This omission is
valid for the case where the final state $|e\rangle$ is stable
enough so that the radiative decay  during the sequence pulses is
negligible. To implement an atom interferometer based on light
pulses the ``clock" transitions between the hyperfine levels of an
atomic ground state constitute an ideal scheme (for instance the
clock transition $|F=3, m_F=0\rangle\longmapsto |F=4, m_F=0\rangle$
in $^{133}$Cs or  $|F=1, m_F=0\rangle\longmapsto|F=2, m_F=0\rangle$
in $^{87}$Rb). In addition, these transitions have no first-order
Zeeman shifts so that the interferometers are insensitive to the
residual magnetic fields. However, since the splitting lies in the
microwave range, the recoil imparted by a single microwave photon
would be only about 0.1 $\mathrm{\mu}$m/s, making the splitting
between the two interfering atomic wavepackets too small to build a
sensitive interferometer. Fortunately, one can obtain a large recoil
and take advantage of the long lifetimes by the use of stimulated
two-photon Raman transitions between these same levels
\cite{Kase91}.
\subsection{Interaction with two counterpropagating waves: stimulated Raman
transition} Let us consider a three level system
(\ket{g},\ket{e},\ket{i}) where two ground state hyperfine levels
\ket{g},\ket{e} are coupled to an intermediate state \ket{i} by two
lasers of angular frequencies $\omega_1$ and $\omega_2$ and wave
vectors $\veck_1$ and $\veck_2$ (see Figure.~\ref{fig:Raman}). The
coupling constants are respectively the Rabi frequencies $\Omega_1$
and $\Omega_2$.

\begin{figure}[h]
\centering
\includegraphics [width=3.5cm]{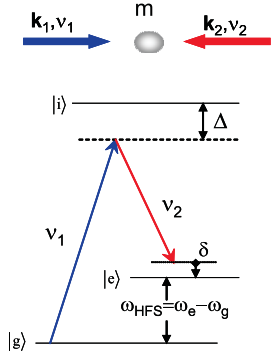}
\caption{\label{fig:Raman} Energy levels and laser frequencies used
for Raman transition.} \label{fig:Raman}
\end{figure}
 In the case where the Raman lasers are far-detuned from resonance with $\ket{g}\longrightarrow\ket{i}$
 or $\ket{e}\longrightarrow \ket{i}$ single
 photon transitions ($\Delta\gg\Gamma_i$ where $\Gamma_i$ is the natural linewidth of the level $\ket{i}$ (see Figure~\ref{fig:Raman})) , the population of the intermediate state \ket{i} remains small and the three level
system can be treated as a two-levels system (\ket{g},\ket{e})
coupled with an effective Rabi frequency $\Omega_{\mathrm{eff}} =
(\Omega_1\Omega_2^*)/2$. The two-photon Raman excitation can be
simply mimed by a single wave of frequency $\omega_1-\omega_2$ and
effective wavevector
$\mathbf{k}_{\mathrm{eff}}=\mathbf{k}_1-\mathbf{k}_2$. This
two-levels system will undergo Rabi oscillations at
$\Omega_{\mathrm{eff}}$ (see the previous paragraph). We notice
first that when the beams are counterpropagating
($\mathbf{k}_1-\mathbf{k}_2\simeq2\mathbf{k}_1$), the transition has
a Doppler sensitivity twice that of a single-photon optical
transition. For $^{87}$Rb atoms, the imparted recoil ($2v_r$) is about
12~mm/s, about five orders of magnitude larger than that of the
corresponding single-photon microwave transition. Second, to
implement successfully stimulated Raman transitions, one does not
require ultrastable laser frequencies : only the difference
$(\omega_1-\omega_2)$ needs to be stable regarding to the hyperfine
transition.

\section{Theoretical treatment of the Ramsey-Bord\'e interferometer}
Atom interferometry based on a light pulses sequence is inspired by
Ramsey's separated oscillatory field methods introduced around 1950
to improve the stability of atomic clocks. Thus the most basic
scheme of atom interferometer is based on a $\pi/2-\pi/2$ pulse
sequence. Later, a pulse sequence of two pairs of $\pi/2$ laser
pulses was first proposed by \cite{Baklanov76,Berquist77} to improve
the resolution in saturation spectroscopy. The interpretation of
four zone Ramsey spectroscopy in terms of atom interferometry with
separated wave packets was suggested by Bord\'e
\cite{Borde89,Borde84}. This scheme has thus become known as a
Ramsey-Bord\'e interferometer.

In this section we analyze such an interferometer. We consider that
each light pulse drives a Raman transition between the two hyperfine
states $|g\rangle$ and $|e\rangle$. We first calculate the phase
shift between interfering paths in the case of a simple
$\pi/2-\pi/2$ sequence. Then  we investigate the different
configurations for implementation of the more relevant
${\pi/2-\pi/2}-{\pi/2-\pi/2}$ scheme. The most general and widely
used approach to calculate the phase shift is based on the path
integral formalism \cite{Borde89,Borde92,Storey94}. However for
simplicity, we shall use an approach based on a plane-wave
decomposition, which is strictly valid only when the
atoms are subjected to uniform forces. This is indeed the case for
the applications we describe.
\subsection{The $\pi/2$-$\pi/2$ interferometer}

\begin{figure}
\begin{minipage}{.5\textwidth}
   \centering
    $a): \pi/2-\pi/2 :$ Ramsey~interferometer\\
  \includegraphics[width=0.9\textwidth]{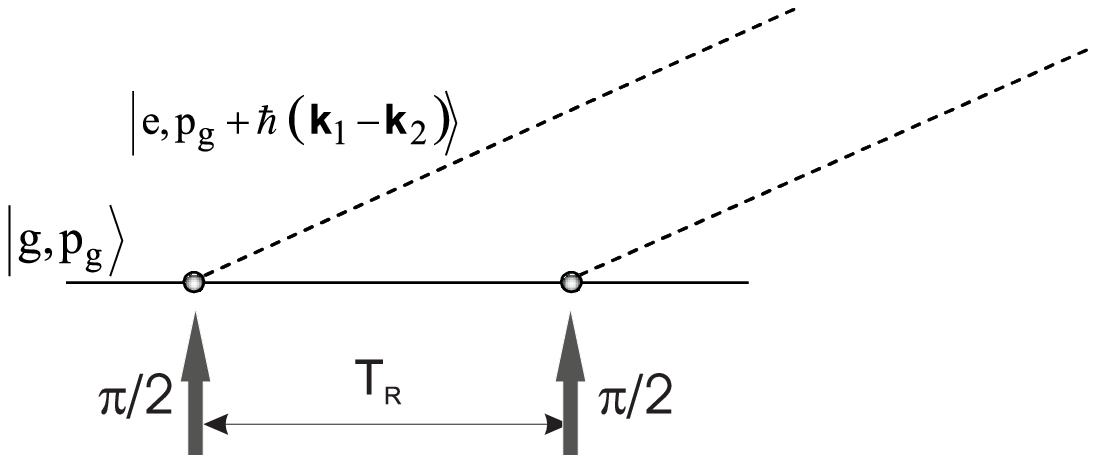}
  \end{minipage}
  \begin{minipage}{.5\textwidth}
   \centering
   $b): \pi/2-\pi-\pi/2:$ Gravimeter\\
  \includegraphics[width=0.9\textwidth]{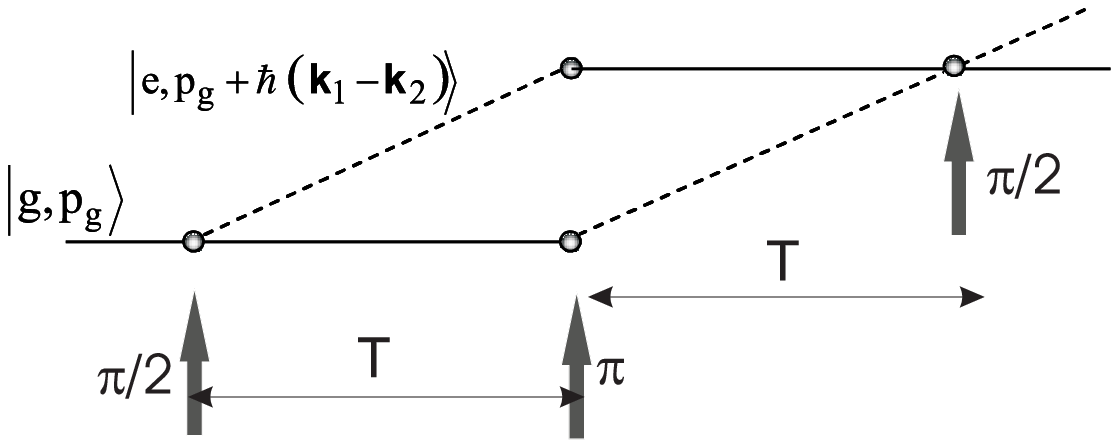}\\
  \end{minipage} \\
  \medskip \\
  \begin{minipage}{.5\textwidth}
     \centering
   c): Symmetric Ramsey Bord\'e interferometer\\
  \includegraphics[width=0.9\textwidth]{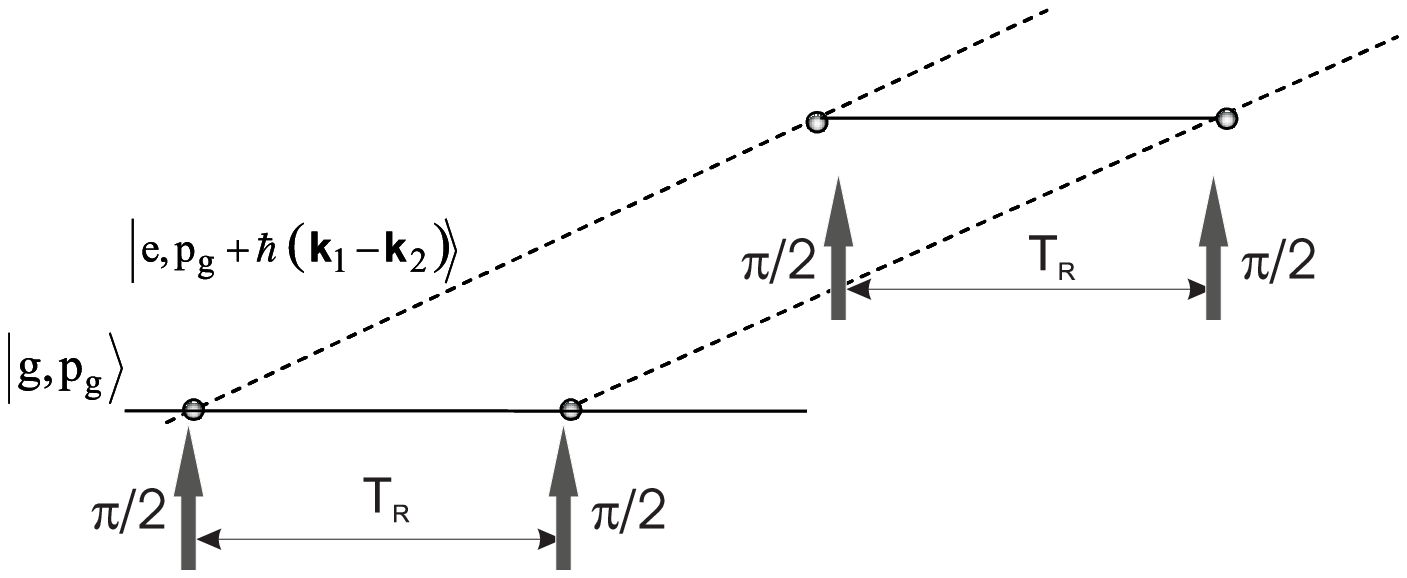}\\
  \end{minipage}
  \begin{minipage}{.5\textwidth}
   \centering
   d): Asymmetric Ramsey Bord\'e interferometer\\
  \includegraphics[width=0.9\textwidth]{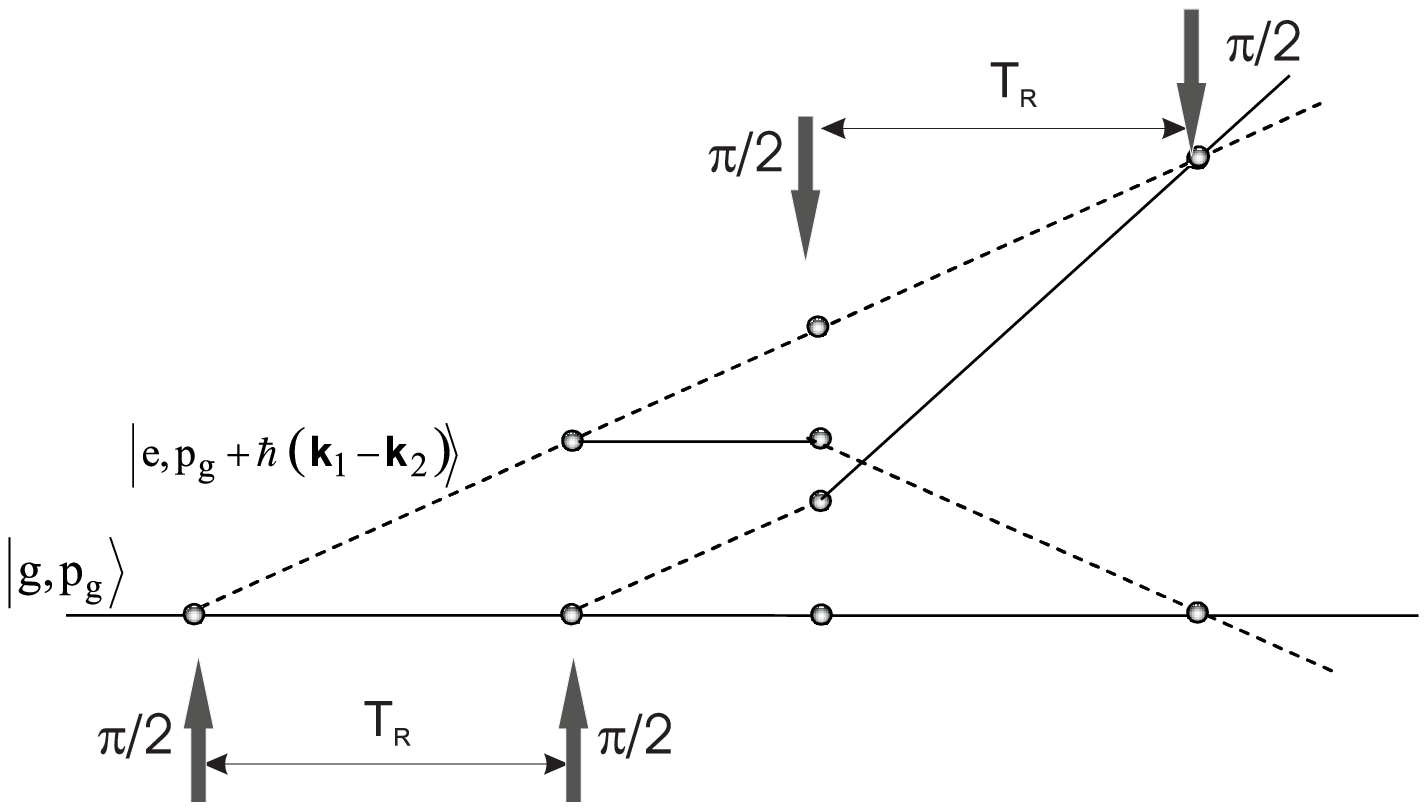}\\
  \end{minipage}
\caption{Recoil diagrams for different configurations of atomic
interferometer based on sequences of light pulses (in time-space domain).}
\label{fig:interferometres}
\end{figure}

For pedagogical purposes, let us first consider the simplest
configuration described in Figure~\ref{fig:interferometres}.a. We
start from an initial single-atom wave plane of momentum $\vecp$ in
the internal state \ket{g}. This state is coupled to the plane wave
of momentum $\vecp + \hbar\veck_1 - \hbar\veck_2$ via a Raman
transition. This transition is induced by two counterpropagating
laser beams $(\mathbf{k}_1, \omega_1)$ and $(\mathbf{k}_2,
\omega_2)$. We denote by $\vecv_g$ the velocity of the atoms in
state \ket{g} and $\vecv_e = \vecv_g + \hbar(\veck_1 - \veck_2)/m$
the velocity of those in state \ket{e}. We apply a sequence of two
$\pi/2$ pulses. After the first  pulse (at time $t_1$), the atoms
are in a coherent superposition: $\ket{\psi} = (\ket{g} -
ie^{-i\phi(t_1)}\ket{e})/\sqrt{2}$. During a time $T_R$ of free
propagation, each state accumulates a phase that depends both on its
internal state and its kinetic energy:
\begin{equation}
\ket{\psi} = \frac{1}{\sqrt{2}}(e^{-i\phi_g}\ket{g} -i
e^{-i(\phi_e+\phi(t_1))}\ket{e})
\end{equation}
where:
\begin{eqnarray}
\phi_g &=& \frac12m\vecv_g^2 \times \frac \TR\hbar \\
\phi_e &=& \frac12m\vecv_e^2 \times \frac \TR\hbar
\end{eqnarray}
After the second $\pi/2$ pulse (at time $t_2$) the atom is in the
following state:
\begin{equation}
\ket{\psi} = \frac{1}{2}\left[(e^{-i\phi_g}
-e^{-i(\phi_e+\phi(t_1)-\phi(t_2)}) \ket{g} -i (e^{-i(\phi_g+
\phi(t_2))}+ e^{-i(\phi_e+ \phi(t_1))})\ket{e}\right].
\end{equation}
The probability to find the atom in the state $\ket{e}$ is then:
\begin{equation}
|\langle e|\psi\rangle|^2=\frac12\left(1+\cos\Phi\right)
\end{equation}
where the phase $\Phi$ is:
\begin{eqnarray}
\Phi = \phi_e - \phi_g+\phi(t_1)-\phi(t_2) &=& \delta \times\TR + \left( \frac12m\vecv_e^2 - \frac12m\vecv_g^2 \right) \times \frac \TR\hbar \\
&=&  \delta\times \TR + \left( \frac{\vecv_e + \vecv_g}2 \right)\cdot
\left( \vecv_e - \vecv_g \right) \frac {m\TR}\hbar
\label{eq:formuleDeuxpulses}
\end{eqnarray}
The key point is that the phase depends both on the mean velocity
and on the velocity difference between the two trajectories. In the
case of counter-propagating Raman transitions, the difference in
velocity is well known and equal to $\hbar (k_1+k_2)/m$. The
interesting quantity is then the mean velocity of the atoms. The
interferometer implemented by two $\pi/2$ light pulses is sensitive
to the initial velocity of atoms. Consequently, the fringe position
will depend on the velocity of the atoms. This means that one can
use this kind of interferometer only in the case where the initial
velocity is well known. This was the case, for example, in the
experiment of Pritchard and co-workers \cite{Campbell}. There, they
started from a Bose Einstein condensate at rest (see Perrin's paper
in this issue). Because the initial velocity distribution was
subrecoil ($\Delta v \ll v_r$), they could assume that $v_g=0$, and
write the phase shift as: $\Phi = \delta \TR + 2\hbar k^2\TR/m$. In
their experiment they were able to measure the recoil energy and
therefore the influence of the refractive index on the recoil
momentum. However, in the case of a wide initial velocity
distribution, the fringes will wash out when one averages over this
distribution. This effect should of course be avoided for
high-precision measurement but is used in some experiments to probe
the initial velocity distribution and coherence properties of gas
\cite{Hugbart2005}. For example, if one starts with a thermal
Maxwell-Boltzmann distribution (velocity distribution in
$e^{-mv^2/(2\textsl{k} T)}$), the phase $\Phi$ has a gaussian
distribution with a variance $\sigma_\mathbf{\Phi}$. Given that the
average value of $\cos\Phi$ is equal to $e^-\frac{\sigma_\Phi^2}2$,
this reduces the contrast of the fringes pattern by a factor of
$\exp\left[-\pi\left(v_r \TR/\lambda_\mathrm{th}\right)^2\right]$
where $\lambda_\mathrm{th} = \sqrt{2\pi \hbar^2/m\textsl{k}T}$ is
the de~Broglie wavelength of an atom of thermal energy $\textsl{k}
T$ ($\textsl{k}$ is Boltzmann constant). This reduction of the
contrast corresponds to the first-order correlation function of a
thermal gas for a distance equal to the separation distance
accumulated between the two $\pi/2$ pulses ($2v_r T_R$).

\subsection{Interferometer with four $\pi/2$ pulses}

To take  full advantage of atom interferometry for high-precision
measurements, one should choose a configuration  where the phase
shift between the two paths is independent of the initial velocity.
This can be achieved if the interferometer is closed \textit{i.e.}
two classical trajectories (the center of mass trajectories)
starting from the same initial point with initial velocity inducing
the two paths of the interferometer end up at the same position
(Figure.~\ref{fig:interferometres}b, \ref{fig:interferometres}c and
\ref{fig:interferometres}d but not \ref{fig:interferometres}a). To
show this in a general case, let us call $v_{1,2}(t)$ the velocity
of center of mass in the first and second arm of the interferometer.
The general formula for the phase shift between the two paths due to
the kinetic energy is:
\begin{equation}
\frac m\hbar\int (\vecv_1(t) -
\vecv_2(t))\cdot\left(\frac{\vecv_1(t) + \vecv_2(t)}2\right) \dt
\label{eq:phasegeneral}
\end{equation}
The velocities $\vec v_1$ and $\vec v_2$ depend linearly on the
initial velocity $v_0$, the phase term depending explicitly on $v_0$
is:
\begin{equation}
\frac{m v_0}{2\hbar}\left(\int  \vecv_1(t) \dt- \int
\vecv_2(t)\dt\right)\
\end{equation}

This term equals zero if $\int\vecv_1(t) \dt = \int \vecv_2(t) \dt$,
\textit{i.e.} the two trajectories are closed.

An obvious scheme that satisfies this requirements is the
$\pi/2-\pi-\pi/2$ sequence. In this case, the first $\pi/2$ pulse
acts as a beamsplitter. The two wave packets are separated by $\hbar
(\mathbf{k_1-k_2})$ (assuming that each pulse drives
Doppler-sensitive Raman transitions). After a time $T$, one applies a
$\pi$ pulse that acts as mirror and redirects the interfering wave
packets so that they overlap at the time $2T$ of the second $\pi/2$
pulse. Closed trajectories can be also obtained by using the
sequence of two pair of $\pi/2$ pulses. The direction of the Raman
beams are reversed for the last pair in order to redirect the atoms.
In the following we discuss the application of such sequences for
the measurement either of the acceleration due to gravity or else
the atomic recoil frequency.

\subsubsection{The $\pi/2-\pi-\pi/2$ scheme : the gravimeter}
The $\pi/2-\pi-\pi/2$ atom interferometer has been first
demonstrated by Kasevich and Chu in 1992 \cite{Kas92} and used for
gravimetry \textit{i.e.} measurement of the acceleration due to
gravity $g$. The light pulses are produced by two
counter-counterpropagating vertical laser beams. Each pulse drives a
Doppler-sensitive Raman transition between two atomic levels. In
this experiment,  Kasevich and Chu changed  the frequency difference
between the two Raman beams $\Delta\omega = \omega_0+\beta (t-t_0)$
in order to compensate the changing velocity of the falling atoms.
If one uses the $\pi/2-\pi-\pi/2$ pulse sequence, the velocity
difference $\vecv_1(t) - \vecv_2(t)$ due to the photon recoil is
equal to $\hbar(\veck_1 - \veck_2)/m$ before the $\pi$ pulse, and
$-\hbar(\veck_1 - \veck_2)/m$ after it. From
eq.~\ref{eq:formuleDeuxpulses} we deduce the total phase :
\begin{eqnarray}
\Phi &=& \beta \TR^2 - \left(\veck_1 - \veck_2\right) \cdot \vecg
\TR^2
\end{eqnarray}

The population in the different internal states is measured scanning
the parameter $\beta$. The value of $\beta$ that leads to $\Phi$=0
is given by:
\begin{equation}
\beta = \left(\veck_1 - \veck_2\right) \cdot \vecg
\end{equation}
In this case , there is no relative phase shift \textit{i.e.} the
changing Doppler shift is exactly canceled by the swept frequency
difference of the Raman laser beams. This condition is insensitive
to the time origin $t_0$ of the sweep and the delay $T_R$ between
the pulses. Note that, in addition to the aforementioned phase
shift, one should take into account the contribution of the initial
net phase of the light beams $\phi^0_i$ at the time $t_i$ when they
are switched on. This leads to an additional phase term
$\phi^0_1-2\phi^0_2+\phi^0_3$ \cite{Peters01}.

\subsubsection{The asymmetric Ramsey-Bord\'e interferometer}
Let us first consider the implementation described in Figure.
\ref{fig:interferometres}d. This kind of interferometer is used to
measure the recoil frequency. Here the directions of the Raman beams
are reversed for the last two $\pi/2$ pulses. Since the difference
of the mean velocity between the first and second part is two
recoils, eq. \ref{eq:formuleDeuxpulses} yields to the following
phase difference between the two arms of the interferometer:
\begin{equation}
\Phi= 2\delta\times \TR + 4\hbar k^2\TR/m
\end{equation}
 In a typical experiment, one varies the detuning $\delta$ in order to record the fringe pattern. The recoil frequency
$\omega_r =\hbar k^2/2m$ is deduced from the central fringe. This
scheme was first used by the Chu group to perform a measurement of
recoil velocity \cite{Chu94,Wicht}.

\subsubsection{The symmetric Ramsey-Bord\'e interferometer}
This interferometer is a generalized version of the gravimeter and
is depicted in Figure. \ref{fig:interferometres}c. It involves a
sequence of four $\pi/2$ pulses without reversal of the direction of
the Raman beams. We assume that the laser beams are oriented
vertically . The delay between each pair of $\pi/2$ pulses is $T_R$
while that between the first and third pulses is $T$. If we consider
that the laser frequencies are swept so that:
\begin{eqnarray}
\delta(t)=\delta_0+\gamma t~~~~~~~~0<t<T_R \\
\delta(t)=\delta_1+\gamma (t -T)~~~~~~~~t>T
\end{eqnarray}
the phase difference between the two arms is then given by
\begin{equation}
\Phi = \TR \left((\delta_1 - \delta_0) +(\vecDeltav + \vecg
T)\cdot(\veck_1 - \veck_2)\right) \label{eq:RBSym}
\end{equation}
where the term $\vecDeltav$ allows for any velocity variation
between the second and third pulses  (on top of the trajectory). The
phase is insensitive to the internal energy of the atoms. As
discussed in the second part of this paper, we have used this scheme
to measure the recoil velocity of atoms. The velocity variation is
induced by the transfer of a large and well-defined number of photon
momenta using a Bloch oscillation phenomena.

\subsection{Description in term of velocity selection}
There is an intuitive way of understanding the Ramsey-Bord\'e
interferometer in terms of selection/measurement of a velocity
distribution. This line of reasoning was adopted by Moler \textit{et
al} in their early paper on atom interferometry \cite{Moler92}.

In the usual Ramsey experiment (as performed for example in atomic
clocks), the transition probability for a system driven with two
$\pi/2$ pulses separated by a delay $\TR$ is a function of the
detuning $\delta$. This function is the product of the sinc function
describing the resonance condition of each $\pi/2$ pulse and a
cosine function describing the interferences fringes, assuming that
$\delta\ll\Omega$ (where $\Omega$ is the effective Rabi frequency):
\begin{equation}
P(\delta) =\frac{\Omega^2}{\Omega^2+\delta^2}
\sin^2\left((\sqrt{\Omega^2+\delta^2})\tau\right)
\cos^2\left(\frac{\delta \times T_R}{2}\right)
\end{equation}
In the case of an atomic velocity sensor, because of the Doppler
effect, the detuning depends on the velocity of each atom. This
means that atoms transferred from state \ket{g} to state
\ket{e} by the first two $\pi/2$ pulses have a velocity distribution
following a Ramsey pattern (velocity comb). The width of each
``tooth" of this Ramsey pattern is proportional to $1/\TR$ while the
position of the central fringe depends on the frequency of the
lasers.

To measure the final velocity distribution, we use a second pair of
$\pi/2$ pulses that transfer atoms from \ket{e} to \ket{g} following
a second Ramsey pattern in velocity space. The proportion of atoms
transferred back to the initial state \ket{g} depends on the
overlaps of the two ``velocity combs". As the frequency of the second
pair of $\pi/2$ pulses is tuned, this relative position varies and
the probability for transferring atoms from \ket{e} to \ket{g}
oscillates. The relative position of the central fringes indicates
the Doppler effect that compensates any velocity change that occurs
between selection and measurement. Furthermore, the longer the delay
between the $\pi/2$ pulses, the narrower is the width of the ``teeth"
of the Ramsey pattern and hence the resolution with which the
frequency of the central peak can be located .

In the case of the symmetric Ramsey-Bord\'e interferometer, without
gravity, if $\delta_0$ (resp. $\delta_1$) are the frequency for the
first (resp. second) pair of $\pi/2$ pulses, then the initial
pattern is centered on the velocity $\vecv_0$ given by $(\veck_2 -
\veck_1)\cdot \vecv_0 = \delta_0$. The value of $\delta_1$
corresponding to the central fringe is then $\delta_0 + (\veck_2 -
\veck_1)\cdot \vecDeltav$. This result is identical to
eq.~\ref{eq:RBSym}.

\section*{\large{Part II : A measurement of $h/m_{\mathrm{Rb}}$ combining atom
interferometer and Bloch oscillations}}
\subsection*{Outline}
\label{sec:1}

Quantum mechanics always links the mass m of a particle to the
Planck constant through the ratio $h/m$. Mass appears in the basic
quantum mechanical equations of motion only in this ratio (in the
Schr\"{o}dinger equation, as well as in the relativistic equations
of Dirac and Klein-Gordon). To compare quantum theories with
experiment, only the measurement of this ratio is required and not $m$
independently. The most relevant example of this is that the ratio
$h/m_e$ ($m_e$ is the electron mass) is involved in the
determination of the fine structure constant $\alpha$, via the
Rydberg constant $R_{\infty}$:
\begin{equation}
\alpha^2=\frac{2 R_{\infty}}{c}\times\frac{h}{m_e}
\end{equation}
The ratio $h/m$ for an atomic system can be deduced from the
measurement of the recoil velocity or energy combined with the
accurate measurement of the photon wavelength $\lambda$. Since the
preliminary measurement of the recoil splitting in saturation
spectroscopy \cite{Hall76}, laser cooling has stimulated renewed
interest in such a determinations \cite{Wicht,Clade06}. The
aim of our experiment is to provide an improved value of the
fine structure constant $\alpha$ from the measurement of the ratio
$h/m$. The two quantities are related via \cite{Taylor}:

\begin{equation}
\alpha^2=\frac{2R_\infty}{c}\frac{A_r(X)}{A_r(e)}\frac{h}{m_X}
\label{alpha-h/m}
\end{equation}
where $R_\infty$ is the Rydberg constant, $A_r(e)$ is the relative
atomic mass of the electron and $A_r(X)$ the relative mass of the
particle $X$ with mass $m_X$. These factors are known with a
relative uncertainty of $7\times 10^{-12}$ for $R_\infty$
\cite{{Udem},{Schwob}}, $4.4\times 10^{-10}$ for $A_r(e)$
\cite{Beir} and less than $2.0\times 10^{-10}$ for $A_r(\rm Cs)$ and
$A_r(\rm Rb)$ \cite{Bradley}. Hence, the factor limiting the
accuracy of $\alpha$ is the ratio $h/m_X$.

The fine structure constant is a corner stone of the adjustment of
the fundamental physical constants \cite{{codata02},{codata06}}. Its
value is obtained from experiments in different domains of physics,
such as the quantum Hall effect in solid state physics, or the
measurement of the muonium ground-state hyperfine structure in
atomic physics (see papers in this issue). At present, the most
precise determinations of $\alpha$ have been deduced from the
measurements of the electron anomaly $a_e$ made in the 1980's at the
University of Washington \cite{VanDick} and, more recently, at
Harvard \cite{{Gabrielse},{Odom}}. This last experiment and an
impressive improvement of the quantum electrodynamics (QED)
calculations \cite{Kino,Gabrielse2007} have led, in 2008, to a new
determination of $\alpha$ with a relative uncertainty of
$3.7\times10^{-10}$ \cite{Gabrielse2008}. Hence, the value of the
fine structure constant recommended by the CODATA is mainly
determined from $a_e$ measurement via QED calculation. This shows
the need for other routes to the value of $\alpha$ independent of
QED. The determination of $\alpha$ deduced from the measurements of
$h/m_{\rm Cs}$ \cite{Wicht} and $h/m_{\rm Rb}$ \cite{Clade06,Clade2}
(where $m_{\rm Cs}$ and $m_{\rm Rb}$ are the masses of cesium and
rubidium atoms) have an uncertainty of $7\times10^{-9}$ and
$6.7\times10^{-9}$ respectively. Nowadays, the method based on the
photon recoil measurement constitutes the unique alternative to
increase our confidence in the recommended value of $\alpha$.

\section{Principle of the experiments}
In our work, the ratio $h/m$ is deduced from the measurement of the
recoil velocity of a $^{87}$Rb atom which absorbs or emits a photon,
and the frequency of the photon involved. We use a symmetric
Ramsey-Bord\'e interferometer (two pairs of $\pi/2$ laser pulses).
Each pulse drives a Doppler-sensitive Raman transition between the
two hyperfine ground states $F=2$ and $F=1$.
\begin{figure}
\centering
\includegraphics [width=10cm]{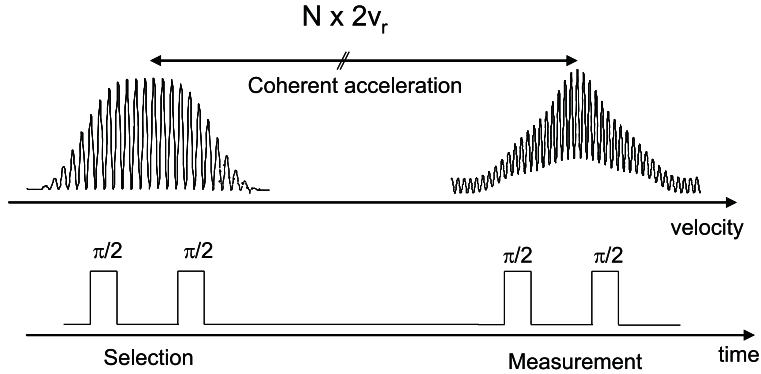}
\caption{We select a narrow velocity ``comb" by a first pair of
$\pi/2$ Raman pulses (transition between the $F=2$ and $F=1$
hyperfine sublevels of the ground state). Then the atoms are
accelerated by means of Bloch oscillations, and the final velocity
of the atoms is probed with a second pair of $\pi/2$-pulses.}
\label{fig:principe}       
\end{figure}
To explain the basic principle of the experiment, we describe our
interferometer in momentum space (see Figure.~\ref{fig:principe}):
First we use a pair of $\pi/2$ pulses to select a ``velocity comb"
in the internal state $F=1$ (corresponding to a Ramsey fringes
pattern). The width of the envelope of this velocity comb varies
inversely with the pulse duration $\tau$, while the fringe
width varies as $1/T_R$. We next transfer to these selected atoms a
large number of photon momenta by means of Bloch oscillations as
explained below. Finally, we use a second pair of $\pi/2$ pulses to
probe  the accelerated velocity combs.  By scanning the frequency of
the last two Raman pulses, we perform the convolution of the initial
velocity comb by the second one. The accuracy of the photon-recoil
measurement depends on the number of recoils ($2N$) we are able to
transfer to the atoms. Specifically, if we locate the center of the
final velocity distribution (central fringe) with an accuracy of
$\sigma_\mathrm{v}$, the accuracy of the recoil velocity measurement
$\sigma_{\mathrm{v_r}}$ is:
 \begin{equation}
\sigma_{\mathrm{v_r}}=\frac{\sigma_\mathrm{v}}{2N}
\label{incertitude}
\end{equation}
\subsection{Bloch oscillation in an accelerated optical lattice}
\begin{figure}
\centering
\includegraphics[width=6cm]{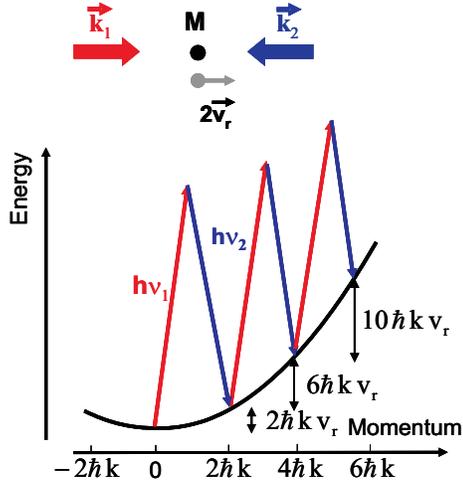}
\caption{Acceleration of cold atoms with a frequency chirped
standing wave. The variation of energy versus momentum in the
laboratory frame is given by a parabola. The energy of the atoms
increases by the quantity $4(2j+1)E_{r}$ at each cycle.}
\label{fig:Parabole}       
\end{figure}
Bloch oscillations were first observed in atomic physics a decade ago,  by groups in Paris \cite{{Dahan},{Peik}}
and in Austin \cite{Wilkinson}. Similar
effects have also been mentioned by Bord\'e \cite{Borde89B}. In a
simple way, they can be viewed as two-photon transitions in which
the atom begins and ends in the same energy level, so that its
internal state ($F=1$) remains unchanged while its velocity has
increased by $2v_r$ per oscillation (see Figure.
\ref{fig:Parabole}). Bloch oscillations are produced in a
one-dimensional optical lattice whose wells are accelerated as the
relative frequency of two counter-propagating laser beams is swept
linearly. The frequency difference $\Delta\nu$ is increased so that,
because of the Doppler effect, the beams are periodically resonant
with the same atoms ($\Delta\nu=4(2j+1)E_{r}/h$, $j=0,1,2,3..$ where
$E_{r}/h$ is the recoil energy in frequency units and $j$ the number
of transitions). This leads to a succession of rapid adiabatic
passages of the atoms between momentum states differing by
$2h\nu/c$. In the solid-state physics approach, this phenomenon is
known as Bloch oscillations in the fundamental energy band of a
periodic optical potential. In an optical lattice, the atoms are
subjected to a constant inertial force obtained by the introduction
of the tunable frequency difference $\Delta\nu$ between the two
waves that create the optical potential \cite{Dahan}. The usefulness
of the Bloch oscillations technique is that it allows one to
transfer a large number of photon momenta (efficiency of $99.97\%$
per recoil) to any velocity shape that lies within the first
Brillouin zone (defined by [-$\hbar k$, +$\hbar k$] in momentum
space), in a coherent way, \textit{i.e.} without any change of the
velocity shape \cite{BattestiPRL03,Clade2}.

Before closing this section, we mention an elegant configuration for
Bloch oscillations, which occurs when an atom is placed in a
vertical standing wave. The atom, initially at rest, starts to fall
because of gravity. When its momentum has reached the value $-\hbar
k$, it absorbs a photon from the up-propagating wave and emits
another one in the down-propagating wave. At the end of this
$\Lambda$ transition, its momentum is equal to $+\hbar k$. The
atomic momentum thus oscillates between $+\hbar k$ and $-\hbar k$.
The time required for this oscillation is equal to $T=2h\nu/cMg$
where $g$ is the acceleration due to gravity, $\nu$ the optical
frequency of the wave and $c$ the speed of the light. In the case of
rubidium atom, the oscillation frequency is about 830~Hz. This
effect has been observed by our group \cite{stat} and briefly described in
previous papers \cite{JPHYS,Battesti03} . It has also been reported by other
groups \cite{{Inguscio},{Tino},{Naegerl}}.

\section{Experimental setup}
\label{sec:2} 
\begin{figure}
\centering
\includegraphics[width=8cm]{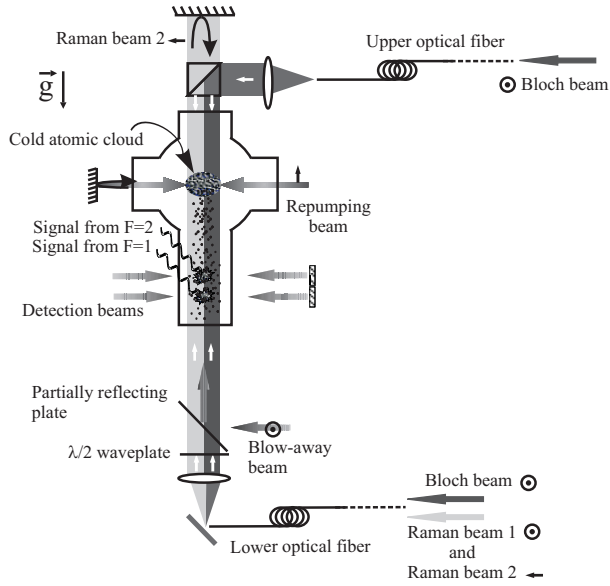}
\caption{Scheme of the experimental setup: the cold atomic cloud is
produced in a MOT (the cooling laser beams are not shown). The Raman
and the Bloch beams are vertical. The Raman beams and the upward
Bloch beam are injected into the same optical fiber. The
``blow-away'' beam is tuned to the one photon transition and allows
us to clear the atoms remaining in $F=2$ after the selection step.
The populations in the hyperfine levels $F=1$ and $F=2$ are detected
by fluorescence 15~cm below the MOT using a time of flight
technique.} \label{fig:beams-setup}
\end{figure}

Our experimental setup (Figure.~\ref{fig:beams-setup})
has been previously described in detail elsewhere \cite{Clade2}.
Briefly, $^{87}\rm Rb$ atoms are captured from a background vapor in
a $\sigma^+-\sigma^-$ configuration magneto-optical trap (MOT). The
trapping magnetic field is then switched off and the atoms are
cooled to about $3~\mu K$ in an optical molasses. After the cooling
process, we apply a bias field of $10~\mathrm{\mu T}$. The atoms
are then optically pumped into the $F=2, m_F=0$ ground state. We
apply the first pair of $\pi/2$ pulses which select a comb of
velocities. The $\pi/2$ pulses are produced by two vertical
counter-propagating laser beams (Raman beams). The selection is
implemented by transferring the resonant atoms from $5 S_{1/2}$,
$\left|F=2, m_F = 0\right>$ state to $5 S_{1/2}$, $\left|F=1,m_F =
0\right>$ state and pushing away the atoms remaining in the state
$F=2$. To push away the unwanted atoms, we apply after the first
$\pi/2-\pi/2$-pulse sequence, a laser beam resonant with the
$5S_{1/2}~(F=2)$ to $5P_{3/2}~(F=3)$ cycling transition. Atoms in
the state $F=1$ make $N$ Bloch oscillations in an accelerated
vertical optical lattice.

We then perform the velocity measurement step using the second pair
of Raman $\pi/2$-pulses, whose frequency is $\delta_{\rm meas}$. The
populations in the levels $F=1$ and $F=2$ are measured separately by
using the one-dimensional time of flight technique. To plot the
final velocity distribution we repeat this procedure by scanning the
  Raman beam frequency $\delta_{\rm meas}$ of the second pulse.

The Raman beams are produced by two frequency-stabilized diode
lasers. Their beat frequency is precisely controlled by a frequency
chain allowing one to switch easily from the selection frequency
($\delta_{\rm sel}$) to the measurement frequency ($\delta_{\rm
meas}$). One of the lasers is locked to a peak of a highly stable
Fabry-Perot cavity. The two Raman beams have linear orthogonal
polarizations and are coupled into the same polarization-maintaining
optical fiber. The pair of Raman beams is sent through the vacuum
cell. The counter-propagating configuration is achieved using a
polarizing beam-splitter cube and a horizontal retro-reflecting
mirror placed above the exit window of the cell.

The standing wave used to create the 1-D optical lattice is
generated by a Ti:Sapphire laser, whose frequency is stabilized to
the same highly stable Fabry-Perot cavity used for the Raman
beam laser diodes. Its frequency is continuously measured by
counting the beatnote with another frequency stabilized Ti-Sapphire laser
referenced to a Rb two-photon  standard \cite{DeuxPhotonRb,Bea00}

\begin{figure}[h]
\centering
\includegraphics[width=8cm]{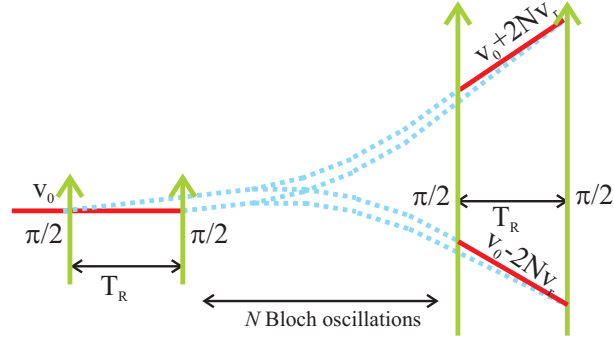}
\caption{Scheme of the interferometers used for the measurement of
$h/m_{\rm Rb}$. The first pair of $\pi/2$ pulses selects a comb of
velocities which is measured by the second pair of $\pi/2$ pulses.
Between these two pairs of pulses, the atoms are accelerated upwards
or downwards. The solid line corresponds to the atom in the $F=2$
level, and the dashed line to the $F=1$ level.}
\label{fig:Interferometre}
\end{figure}

The Ti:Sapphire laser beam is split into two parts. To implement the
timing sequence, we use several acousto-optic modulators to control the
intensity and frequency of different laser beams. The optical
lattice is blue detuned by $\sim 40~\mathrm{GHz}$ from the
$5S_{1/2}-5P_{3/2}$ single photon transition. The power of the
lattice beams is raised slowly (500~$\mu$s) in order to load all the
atoms adiabatically into the first Bloch band. To perform the
coherent acceleration, the frequency difference of the two laser
beams generating the optical lattice is swept linearly. Then, the
lattice intensity is slowly lowered in 500~$\mu$s to bring atoms
back into a well-defined momentum state. The optical potential depth
is $\sim 100~E_r$. With these lattice parameters, the spontaneous
emission is negligible. For an acceleration of about $2000~\rm m\rm
s^{-2}$ we transfer about 1200 photon momenta in 3~ms. To avoid
atoms from reaching the upper window of the vacuum chamber, we use a
double acceleration scheme: instead of selecting atoms at rest, we
first accelerate them using Bloch oscillations and then we perform
the three-step sequence: selection-acceleration-measurement. In this
way, the atomic velocity at the measurement step is close to zero.
In the vertical direction, an accurate determination of the recoil
velocity would require an accurate measurement of the acceleration
due to  gravity $g$. In order to circumvent this difficulty, we
make a differential measurement by accelerating the atoms in
opposite directions (upward and downward trajectories) keeping the
same delay between the two pairs $\pi/2$-pulses. Thus the
photon-recoil measurement is a determination of the frequency
difference between the central fringes of two opposite
interferometers (upward and downward) illustrated in
Figure.~\ref{fig:Interferometre}.

The ratio $\hbar/m$ can then be deduced from the formula:
\begin{equation}
\frac{\hbar}{m}= \frac{(\delta_{\rm sel}-\delta_{\rm meas})^{\rm up}
- (\delta_{\rm sel}-\delta_{\rm meas})^{\rm down}}{2(N^{\rm
up}+N^{\rm down})k_B (k_1+k_2)} \label{eq:mynum}
\end{equation}
where $(\delta_{\rm meas}-\delta_{\rm sel})^{\rm up/down}$
corresponds respectively to the center of the final velocity
distribution for the up and the down trajectories, $N^{\rm up/down}$
are the number of Bloch oscillations in both opposite directions,
$k_B$ is the Bloch wave vector and $k_1$ and $k_2$ are the wave
vectors of the Raman beams.

Moreover, the contribution of some systematic effects (energy level
shifts) to $\delta_{\rm sel}$ and $\delta_{\rm meas}$ changes sign
when the directions of the Raman beams are interchanged. To improve
the experimental protocol, for each up or down trajectory, the
directions Raman beams are reversed and we record two velocity
spectra. Finally, each determination of  $h/m_{\rm Rb}$  and
$\alpha$ is obtained from four velocity spectra.

\begin{figure}[h]
\begin{minipage}{.5\textwidth}
   \centering
  \includegraphics[width=0.8\textwidth]{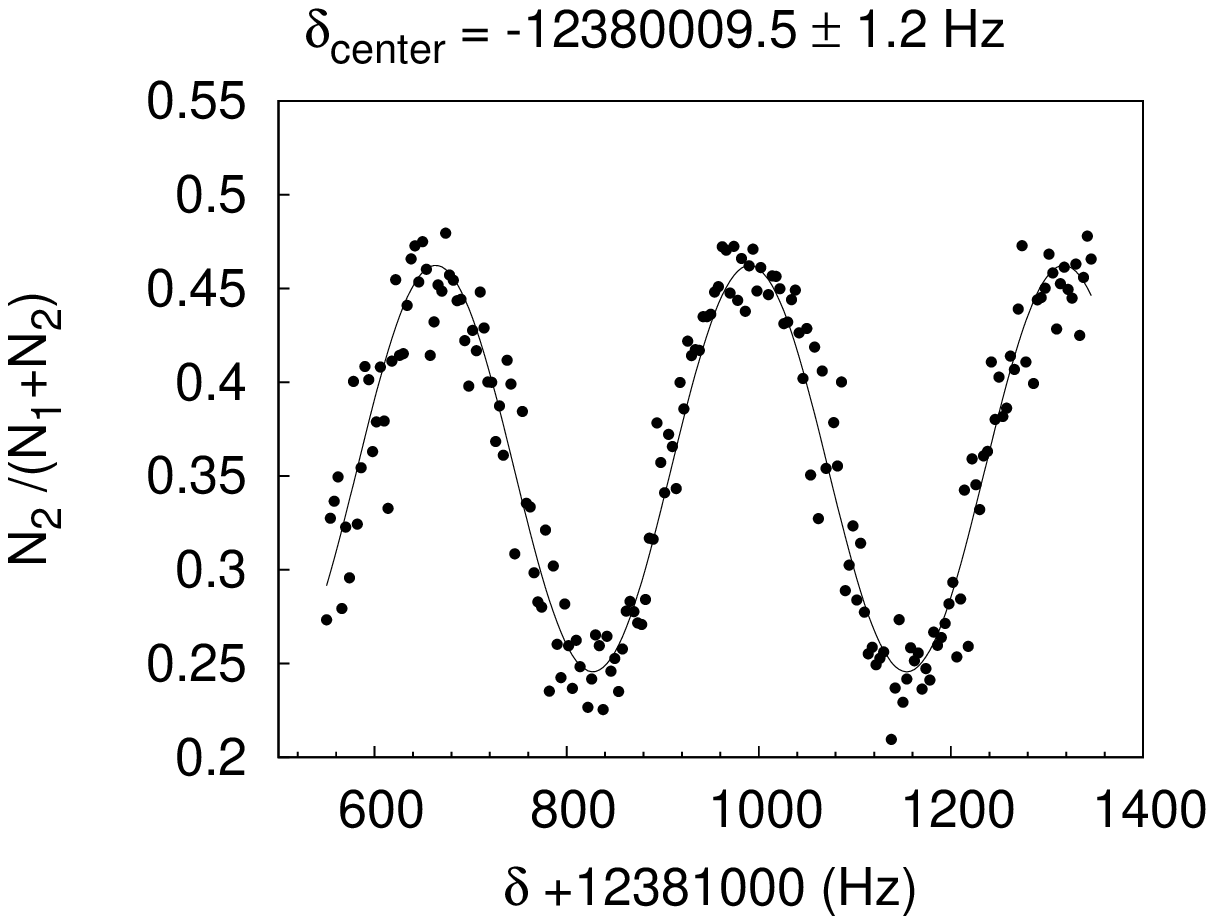}
  \end{minipage}
  \begin{minipage}{.5\textwidth}
   \centering
  \includegraphics[width=0.8\textwidth]{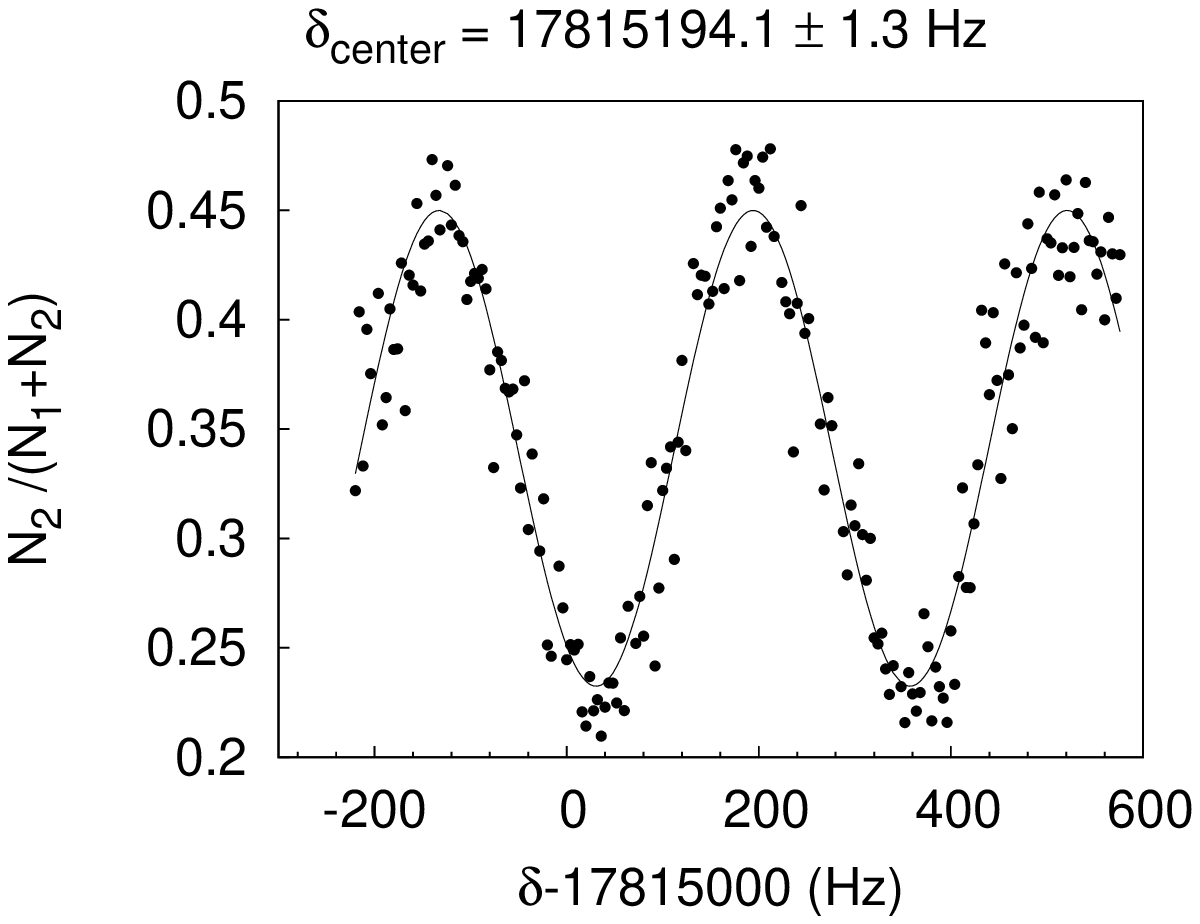}\\
  \end{minipage} \\
  \medskip \\
  \begin{minipage}{.5\textwidth}
     \centering
  \includegraphics[width=0.8\textwidth]{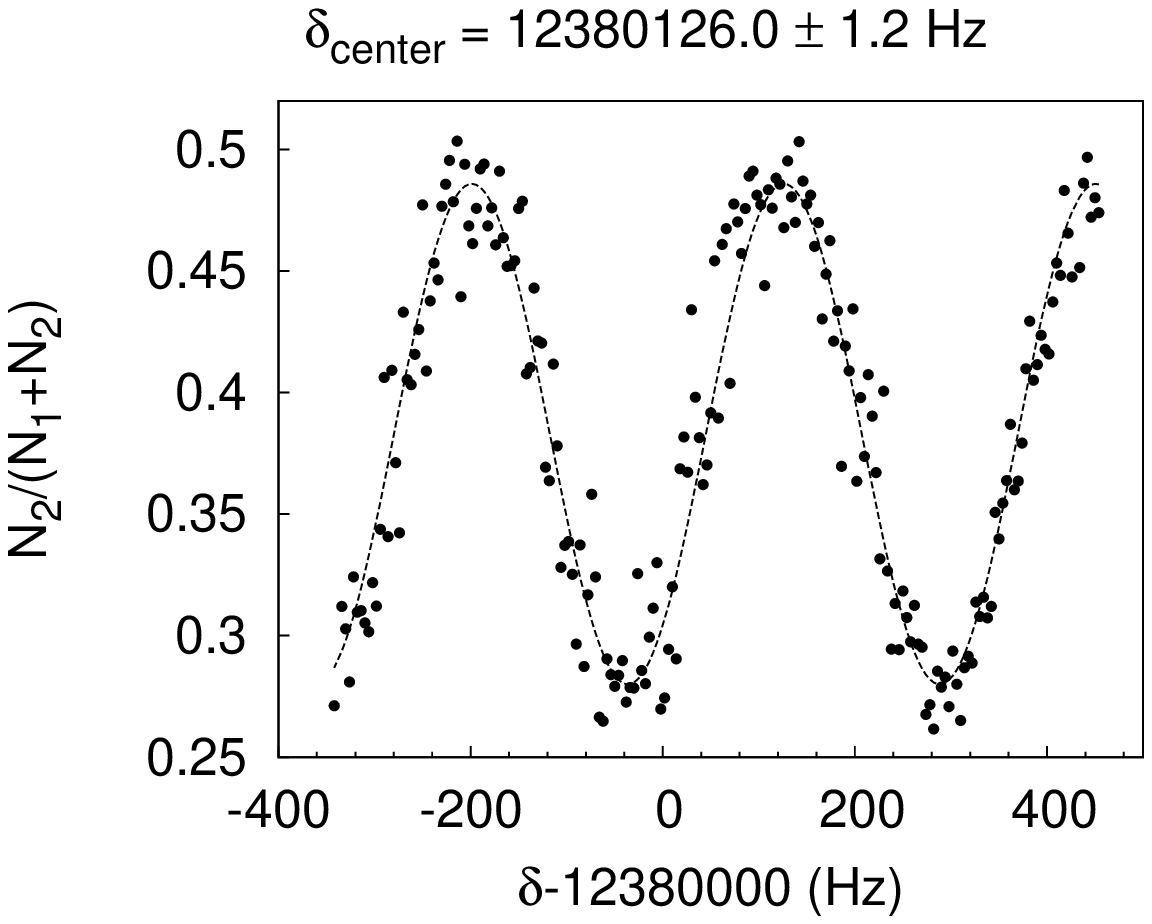}\\
  \end{minipage}
  \begin{minipage}{.5\textwidth}
   \centering
  \includegraphics[width=0.8\textwidth]{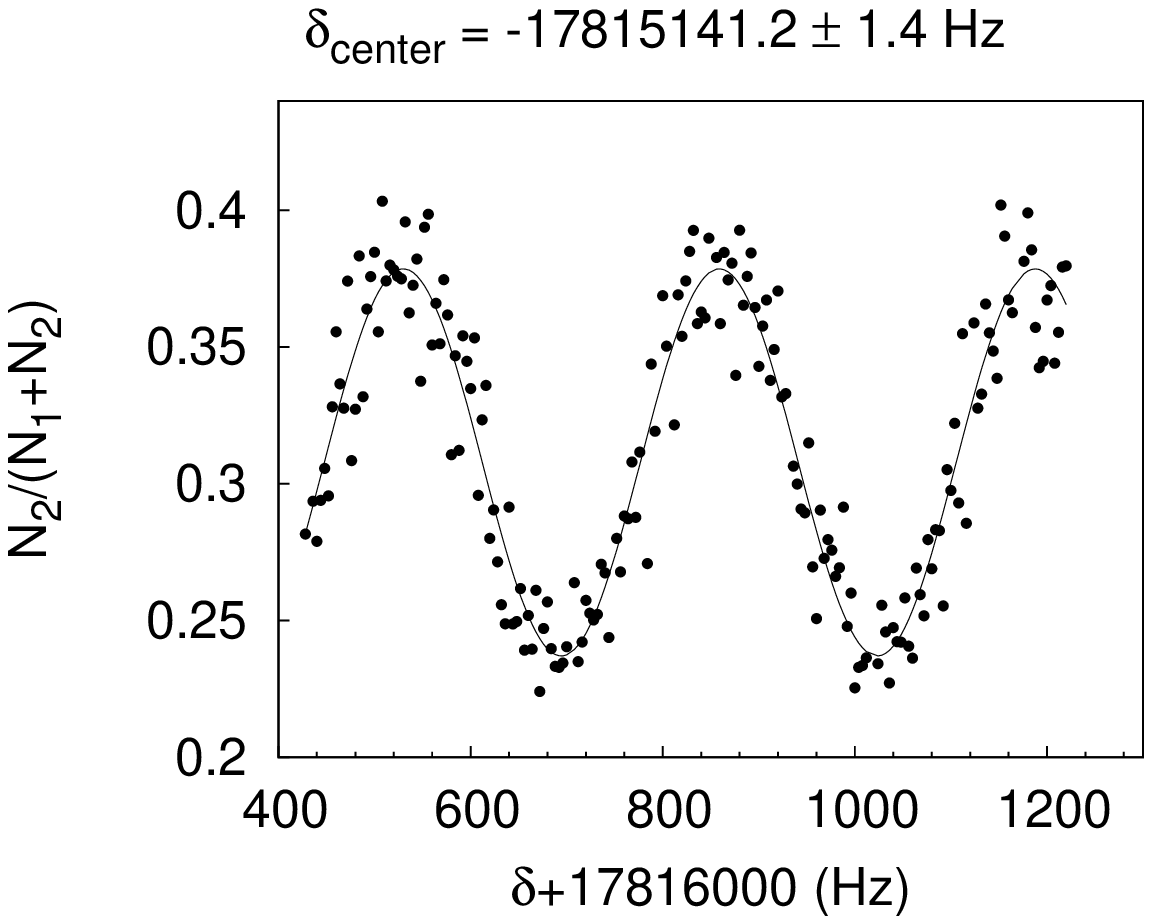}\\
  \end{minipage}
\caption{Spectra obtained by atomic interferometry which show the
proportion of atoms in the $F=2$ level ($N_2/(N_1+N_2)$) in function
of the frequency difference $\delta_{\rm meas}-\delta_{\rm sel}$ in
hertz. The two spectra on the left corresponds to the upwards
acceleration (400 Bloch oscillations) and the two spectra on the
right to the downwards acceleration (600 Bloch oscillations).  }
\label{fig:quatre-spectre2}
\end{figure}
\section{Preliminary results}
\label{sec:4} In this section we present our recent and preliminary
determination of the fine structure constant. The measurement of the
ratio $h/m_{\mathrm{Rb}}$ is obtained by combining a Ramsey-Bord\'e
interferometer with Bloch oscillations in an accelerated lattice
(see the experimental protocol described in
Figure.~\ref{fig:Interferometre}). A typical set of four spectra
used to provide one determination of the ratio $h/m_{\mathrm{Rb}}$
is shown in Figure.~\ref{fig:quatre-spectre2}. In this case, the
total number of Bloch oscillations is $N^{\rm up}+N^{\rm
down}=1000$, corresponding to 2000 recoil velocities between the up
and down trajectories. The duration of each $\pi/2$ pulse is 400
$\mu$s and the time $T_R$ is 2.6 ms (the total time for a pair of
$\pi/2$ pulses is 3.4 ms). The central fringe is determined with an
uncertainty lower than $1.4$~Hz ($\sim v_r/10000$). Each spectrum is
plotted with 200 points and is obtained in $6$~min. We obtain an
excellent fringes visibility of about 30$\%$ for 600 Bloch
oscillations. The analysis of 125  determinations of $\alpha$ leads
to the  statistical uncertainty of $3.3\times 10^{-9}$ with
$\chi^2/(n-1)$=1.55.

The systematic effects taken into account to determine the value of the fine structure constant are summarized in Table \ref{Systematic}. These effects are detailed in a previous paper \cite{Clade2}. Up to now, the main contributions come from the corrections due to the wavefront curvature
and the second-order Zeeman effect due to the residual magnetic field. This last correction is determined by  carefully mapping
the magnetic field in the interaction area. It should be reduced by implementing a magnetic shielding.
The total relative uncertainty on $\alpha$ is
then $4.7 \times 10^{-9}$.

The preliminary value of the fine structure constant is:

\begin{equation}
\alpha^{-1}=
137.035\,998\,87\,(64)~~~~[4.7\times10^{-9}]
\end{equation}
This value is in good agreement with our last determination $\alpha [06]$ published in 2006 \cite{Clade2}. It was obtained by using an non-interferometric method based on the $\pi-\pi$ Raman pulses for the velocity sensor.

\begin{equation}
\alpha^{-1}[06]= 137.035\,998\,84\,(91)~~~~[6.7\times10^{-9}]
\end{equation}

Complementary measurements are in progress in order to determine the final value of the fine structure constant based on this interferometric approach.

\begin{table}
\begin{center}
\begin{tabular}{lcc}
\hline
\multicolumn{1}{l}{Source}
&Correction  &Relative uncertainty \\
&($\times10^{-9}$)  &($\times10^{-9}$)\\
\hline
 Laser frequencies& &0.4\\
Beams alignment&-2& 2\\
Wavefront curvature and Gouy phase&-11.9 & 2.5\\
2nd order Zeeman effect&7 & 1 \\
Quadratic magnetic force&-1.45 & 0.2\\
Gravity gradient&-0.18& $0.02$ \\
light shift (one photon transition)& 0& 0.1\\
light shift (two photon transition)&0 & 0.01 \\
light shift (Bloch oscillation)&0.58 & 0.2 \\
Index of refraction atomic cloud& 0&0.3 \\
Index of refraction background vapor&-0.41 & 0.3 \\
\hline \hline
Global systematic effects&-8.36 & 3.4\\
\hline
\end{tabular}

\caption{\label{Systematic} Error budget on the determination of
$\alpha$ (Systematic effects and relative uncertainty). }
\end{center}
\end{table}

\section{Conclusion}
\label{sec:5}

In this lecture we have given a brief analysis of atom
interferometers based on the arrangements of $\pi/2$ and $\pi$ laser
pulses. We have discussed their applications to the measurement of
the gravitational acceleration and atomic recoil frequency. We have
described in detail an experimental method which combines the Bloch
oscillation process with a Ramsey-Bord\'e interferometer. We believe
that this combination is a very promising method for improving the
accuracy of atomic recoil measurements.

An alternative way to improve the sensitivity of atom interferometer
consists in using a large momentum beam-splitter.
 Such beam-splitters can
be implemented using multi-photon Bragg diffraction in a static
optical lattice \cite{Borde89,Giltner95,Gupta02,Koolen02} or Bloch
oscillations in a moving lattice \cite{Denshlag02}. In this case,
the separation between the interfering atomic wavepackets is
performed by 2$N \hbar k$ instead $2\hbar k$ for a $\pi/2$ light
pulse (where $N$ is the number of photon momenta imparted to the
atoms as the result of the interaction with the optical lattice). In
this way, one increases the area of the interferometer and thus its
sensitivity. These investigations are in progress \cite{Chu07} and
should lead to a growing number of applications of atom
interferometry, metrology and applied physics.

\vspace{1cm}
 This experiment is supported in part by the Laboratoire
National de M\'etrologie et d'Essais (Ex. Bureau National de
M\'etrologie)(Contrat 033006), by IFRAF (Institut Francilien de
Recherches sur les Atomes Froids) and by the Agence Nationale pour
la Recherche, FISCOM Project-(ANR-06-BLAN-0192).

\end{document}